\documentclass[twocolumn,showpacs,pra,aps,superscriptaddress,amssymb]{revtex4-1}
\usepackage{amsfonts}
\usepackage{mathrsfs}
\usepackage{amssymb}
\usepackage{bm}
\usepackage{graphicx}
\usepackage[colorlinks=true,linkcolor=blue,citecolor=blue,urlcolor=blue]{hyperref}

\begin{document}
\title{Underlying non-Hermitian character of the Born rule in open quantum systems}
\author{Gast\'on Garc\'{\i}a-Calder\'on}
\email{gaston@fisica.unam.mx}
\affiliation{Instituto de F\'{\i}sica, Universidad Nacional Aut\'onoma de M\'exico,
Apartado Postal {20 364}, 01000 M\'exico, Ciudad de M\'exico, Mexico}
\author{Lorea Chaos-Cador}
\email{lorea@ciencias.unam.mx}
\affiliation{Universidad Aut\'onoma de la Ciudad de M\'exico,
Prolongaci\'on San Isidro 151, 09790 M\'exico, Ciudad de M\'exico, Mexico}

\begin{abstract}
The absolute value squared of the probability amplitude  corresponding to the overlap of an initial state with  a continuum wave solution to the Schr\"odinger equation of the problem, has the physical interpretation provided by the Born rule. Here, it is shown that for an open quantum system, the above probability  may be written in an exact analytical fashion as an expansion in terms of  the non-Hermitian resonance (quasinormal) states and complex poles to the problem which provides an underlying non-Hermitian character of the Born rule.
\end{abstract}
%
%\shortabstract
%
%\begin{document}

%\pacs {03.65.Ca,03.65.Db}

\maketitle

\section{Introduction}
It is well known that Max Born stated the rule that now bears his name in the early days of quantum mechanics \cite{born26}. This rule provides a probabilistic interpretation for the wave function and establishes a link between the formalism of quantum mechanics and experiment, but more importantly, it represents the irruption of indeterminism in the description of matter at microscopic scale. This interpretation has had a profound impact into the notion of what is real and has initiated a debate that remains alive up to the present day as shown by distinct interpretations of quantum mechanics \cite{auletta00,laloe01,kastner12,cetto14}, of studies on the classical-quantum transition \cite{schlosshauer07}, studies concerned with  topics as the reality of the wave function \cite{spekkens07,ringbauer15,barrett16}, and on the Born rule \cite{landsman09,towler12}.

Here we refer to the Born rule for open quantum systems characterized by a continuous spectrum. For that purpose we consider a simple problem, namely, the time honored problem of a particle confined initially within a finite region of space by a potential from which it escapes to the outside by tunneling. The time-dependent wave function may be expanded in terms of the \emph{continuum wave functions} to the problem, and, as is well known, the absolute value squared of the probability amplitude  corresponding to the overlap of the corresponding initial state with a continuum wave solution to the Schr\"odinger equation to the problem, has the physical interpretation provided by the Born rule \cite{griffiths05,cohen05}. The above probability involves an integral over values of the momentum that extends from zero to infinity, and constitutes a `black box' type of numerical calculation from which little physical insight may be obtained.

In this work we derive an exact analytical expression for the above probability by expanding the corresponding \emph{continuum wave solutions} in terms  of \emph{resonance (quasinormal) states} that we believe provides a deeper physical insight on the Born rule. These states may be defined from the residues at the complex poles of the outgoing Green's function to the problem and allow for an exact analytical non-Hermitian formulation of the description of decay by tunneling in open quantum systems \cite{gc10,gc11}.

In fact, the present work has been motivated by the recent result that the time evolution of decay by tunneling involving \emph{continuum wave functions} yields identical results to that using \emph{resonance (quasinormal) states} \cite{gcmv12}.  The reason is that both basis follow from the analytical properties of the Green's function to the problem. However, whereas the physical meaning of the expansion coefficients involving \emph{continuum wave functions} is provided by the Born rule, that does not occur in the case of \emph{resonance (quasinormal) states}.

We find of interest to elaborate a little bit on the notion of open quantum system that we employ here. If, as time evolves, a particle initially confined within a region of space cannot escape to the outside, the system is said to be a closed system. In that case, the system possesses a purely discrete energy spectrum and exhibits unitary time evolution. On the contrary, if the particle can escape by tunneling to the outside, the system constitutes an open system that has the distinctive feature of exhibiting a continuous energy spectrum. Since energy can escape to the outside the time evolution is non unitary. However, the  continuity equation is fulfilled and hence the total flux is conserved.

Our approach considers the full Hamiltonian $H$ to the problem and relies on the analytical properties of the outgoing Green's function
in the complex momentum plane. It is worth mentioning that there are approaches where the full Hamiltonian $H$ to the system is separated into a part $H_0$, corresponding to a closed system, and a part $H_1$ which couples the closed system to the continuum. This is usually treated to some order of perturbation. This type of approximate approaches has become a standard procedure in the treatment of open systems where perturbation theory can be justified. It has its roots in the old work of Weisskopf and Wigner \cite{wigner30b}. There are also related approaches, that have a great deal of attention nowdays because of its implications for quantum information theory, which are referred to  in the literature also as open quantum systems. Here, a quantum system $S$ is coupled to another quantum system $O$ called the environment and hence it represents a subsystem of the total system $S+O$. Usually it is assumed that the combined system is closed, however due to the interactions with the environment the dynamics of the subsystem $S$ does not preserves unitarity and hence it is referred to as an open system. These approaches involve many degrees of freedom and the description of the mixed states of the total system is made in terms of the density matrix \cite{breuer10,huelga12}. The case where the environment may be neglected and still energy can escape from the system $S$ in a non-perturbative fashion would essentially correspond to the notion of open system considered here.

The paper is organized as follows. In section 2 we briefly review the main aspects of the time evolution of decay using \emph{continuum wave functions} and \emph{resonance (quasinormal) states} and refer to the Born rule. In section 3 we derive the expansion of the coefficient of the wave solution in terms of resonance (quasinormal) states. In Section 4, we illustrate our findings by considering an exactly solvable model, and  finally, Section 5 deals with some concluding remarks.

\section{Time-dependent solution}

Let us consider, to bring the problem into perspective, the time evolution of decay of a particle that is initially confined by a real spherical potential of arbitrary shape in three dimensions. Without loss of generality we restrict the discussion to $s$ waves. Notice that the description holds also  on the half-line in one dimension. We consider an interaction potential of arbitrary shape $ V(r)$ that vanishes after a finite distance, \textit{i.e.} $V(r)=0$ for $r > a$. This is justified on physical grounds for a large class of systems, in particular artificial quantum systems as double-barrier resonant structures \cite{sollner83} or ultracold atoms \cite{jochim11}. Also, since we are interested in the continuum, we refer to potentials that do not hold bound states. The units employed are $\hbar=2m=1$.

The solution to the time-dependent Schr\"odinger equation in the radial variable $r$, as an initial
value problem, may be written at time $t > 0$ as
\begin{equation}
\Psi(r,t)=\int_0^a g(r,r';t)\Psi(r',0) \,dr',\quad t >0\,,
\label{e1}
\end{equation}
where $g(r,r';t)$ stands for the retarded time-dependent Green's function, which may be written in terms of the outgoing Green's function as,
\begin{equation}
g(r,r';t)=\frac{i}{2\pi}\int_{-\infty}^\infty G^+(r,r';k) {\rm e}^{-ik^2t}\,2kdk, \quad t> 0\,.
\label{e2}
\end{equation}
\subsection{The time-dependent solution in terms of  continuum wave functions and the Born rule}

Equation (\ref{e2}) may be used to derive the well known expression of the time-dependent wave evolution
in terms of the \emph{continuum wave solutions} $\psi^+(k,r)$ \cite{newtonchap12},
\begin{equation}
\Psi(r,t)=\int_0^{\infty} C(k)\psi^{+}(k,r) {\rm e}^{-i k^2t}\,dk\,,
\label{d34}
\end{equation}
where the expansion coefficient $C(k)$ is given by
\begin{equation}
C(k)=\int_0^a [\psi^+(k,r')]^*\Psi(r',0)\,dr'\,.
\label{d35}
\end{equation}
The \emph{continuum wave functions} $\psi^{+}(k,r)$ are solutions to the Schr\"odinger equation of the problem
\begin{equation}
[k^2-H] \psi^+(k,r)=0\,,
\label{d36}
\end{equation}
satisfying \cite{newtonchap12},
\begin{eqnarray}
&&\psi^+(k,0)=0
\label{d37a} \\ [.3cm]
&&\psi^+(k,r)=\sqrt{\frac{2}{\pi}}
\frac{i}{2}\left [{\rm e}^{-ikr}-\textbf{S}(k){\rm e}^{ikr} \right ] \quad r \geq a\,,
\label{d37b}
\end{eqnarray}
where $\textbf{S}(k)$ is the \textbf{S}-matrix of the problem.The factor $\sqrt{2/\pi}$ in Eq. (\ref{d37b}) arises from the Dirac delta normalization.

Using Eq. (\ref{d34}) one may calculate, in particular, two quantities that are of interest in decay problems. One of them is the survival amplitude $A(t)$, which gives the probability amplitude that at time $t$ the decaying particle is still described by the initial state $\Psi(r,0)$, namely,
\begin{equation}
A(t)=\int_{0}^{a}\Psi^*(r,0) \Psi(r,t)\,dr\,.
\label{defa}
\end{equation}
Notice that if the initial state $\Psi(r,0)$  is normalized to unity, then $A(0)=1$, which is a probability. Substitution of Eq. (\ref{d34}) into Eq. (\ref{defa}), using (\ref{d35}), gives
\begin{equation}
A(t)=\int_{0}^{\infty}|C(k)|^2 {\rm e}^{-ik^{2}t}\,dk,
\label{11d}
\end{equation}
which corresponds to a probability amplitude due to the effect of the time evolving factor $\exp(-ik^2t)$. The survival probability is defined as $S(t)=|A(t)|^2$. The other quantity of interest is the non-escape probability $P(t)$, that provides the probability that at time $t$, the decaying particle is found within the the interaction region $r<a$,
\begin{eqnarray}
&&P(t)=\int_0^a\, |\Psi(r,t)|^2\,dr= \nonumber \\ [.3cm]
&&\int_{0}^{\infty} dk\,'\int_{0}^{\infty}dk\,\, C^*(k\,')C(k)\times \nonumber \\ [.3cm]
&&\int_0^a dr[\psi^+(k\,',r)]^*\psi^+(k,r)\, {\rm e}^{-i(k^2 -k'\,^2)t}\,.
\label{12d}
\end{eqnarray}
The Eqs. (\ref{11d}) and (\ref{12d}), depend on the expansion coefficient $C(k)$ which, therefore, plays a relevant role in studies on quantum decay involving the basis of \emph{continuum wave functions}.

It is well known, that the Born rule establishes that the probability density for a measurement of the momentum $k$ gives a result in the range $dk$ is \cite{griffiths05,cohen05}
\begin{equation}
d\mathcal{P}(k) = |C(k)|^2\,dk\,,
\label{a1}
\end{equation}
where $C(k)$ is given by Eq. (\ref{d35}). According to the usual interpretation of quantum mechanics, upon measurement the wave function $\Psi(r,0)$ ``collapses'' around a narrow range of \emph{continuum wave functions} $\psi^+(k,r)$ of the measured value. If the initial state is normalized to unity, it then follows that
\begin{equation}
\int_0^\infty \,|C(k)|^2\,dk=1\,.
\label{a2}
\end{equation}
\subsection{Relationship between $\psi^+(k,r)$ and the outgoing Green's function}

The analytical properties of the outgoing Green's function to the problem are the relevant quantity  in the derivation of Eq. (\ref{d34}). This expression is given by an integral that involves only real values of the momentum $k$ \cite{newtonchap12}.

The outgoing Green's function obeys the equation,
\begin{equation}
[k^2-H]G^+(r,r';k)=\delta(r-r')\,,
\label{d11}
\end{equation}
with boundary conditions,
\begin{equation}
G^+(0,r';k)=0;\quad
\left [\frac{\partial}{\partial r}G^+(r,r';k) \right ]_{r=a}=ik G^+(a,r';k)\,.
\label{a3}
\end{equation}

Using the Green's theorem between Eqs. (\ref{d36}) and (\ref{d11}) together with the conditions given by Eqs.
(\ref{d37a}), (\ref{d37b}) and (\ref{a3}) yields \cite{gc76},
\begin{equation}
\psi^{+}(k,r)=-\sqrt{\frac{2}{\pi}}kG^+(r,a;k){\rm e}^{-ika},\quad r \leq a\,.
\label{n2}
\end{equation}
The above expression relates, for a given value of the momentum $k$,  the \emph{continuum wave function}
with the outgoing Green's function to the problem along the internal interaction region. As shown below, Eq. (\ref{n2}) constitutes in our analysis the relevant expression to relate the \emph{continuum wave functions} with the the \emph{resonance (quasinormal) states} of the system.

It is of interest to recall that the outgoing Green's function  may be written in terms of the  \emph{regular}, $\phi(k,r)$, and \emph{irregular}, $f_{\pm}(k,r)$, solutions to the Schr\"odinger equation, obeying respectively, boundary conditions: $\phi(k,0)=0$, $ [{\rm d}\phi(k,r)/{\rm d}r]_{r=0}=1$ and $f_{\pm}(k,r)={\rm e}^{\pm ikr}$ for $r \geq a$, and the Jost function $J_{\pm}(k)=f_{\pm}(k,0)$ as \cite{newtonchap12},
\begin{equation}
G^+(r,r';k)=-\frac{\phi(k,r_<)f_+(k,r_>)}{J_+(k)}\,,
\label{d13}
\end{equation}
where $r_<$ and $r_>$ stand respectively for the smaller and larger of $r$ and $r'$.
The functions $f_{\pm}(k,r)$ are linearly independent and hence $\phi(k,r)$ may be written as
\begin{equation}
\phi(k,r)= \frac{1}{2ik}\left [ J_{-}(k)f_{+}(k,r)-J_{+}(k)f_{-}(k,r) \right ]\,.
\label{e4}
\end{equation}
The \emph{continuum wave functions} may also be written as \cite{newtonchap12}
\begin{equation}
\psi^{+}(k,r)= \sqrt{\frac{2}{\pi}}\frac{k\phi(k,r)}{J_{+}(k)}\,.
\label{e4a}
\end{equation}
In fact, using Eq. (\ref{e4}) into (\ref{e4a}) yields, for $ r\geq a$, Eq. (\ref{d37b})  with
$\textbf{S}(k)=J_{-}(k)/J_{+}(k)$. Since flux conservation requires that $\textbf{S}(k)\textbf{S}^{*}(k)=1$, it follows that $J_{-}^{*}(k)= J_{+}(k)$.

\subsection{Relationship between $G^+(r,r^\prime;k)$ and resonance (quasinormal) states}

The expression for the outgoing Green's function given by Eq. (\ref{d13}) has been used to study in a rigorous form its analytical properties away from real values of $k$ into the complex momentum plane \cite{newtonchap12}. For potentials of arbitrary shape  vanishing exactly after a distance, as considered here, the function $G^+(r,r^\prime;k)$ may be extended analytically to the entire complex $k$ plane, where it has an infinite number of poles, distributed in a well known fashion, corresponding to the zeros of the Jost function $J_{+}(k)$. In fact, a finite number of them lie on  the positive and the negative imaginary $k$-axis, corresponding respectively to bound and antibound states, and the rest, an infinite number of poles, are located in the lower half of the $k$ plane, where due to time-reversal considerations, they are distributed symmetrically with respect to the imaginary $k$-axis. Thus, for  pole at  $\kappa_n=\alpha_n-i \beta_n$ located on the fourth quadrant of the $k$ plane, there corresponds a pole $\kappa_{-n}= -\kappa^*_n$.  As discussed below, these poles correspond to the resonant (quasinormal) states of the problem. In fact, all  states arise from the residues of the outgoing Green's function at these poles.

The residues at the poles $\kappa_n$ of the outging Green's function $G^+(r,r';k)$ are proportional to the \emph{resonance (quasinormal) states} to the problem. They may be obtain, as discussed in Ref. \cite{gcr97},  by adapting to the $k$ plane the derivation in the energy plane given in Ref. \cite{gcp76}, namely,
\begin{equation}
\rho_n(r,r')= \frac{u_n(r) u_n(r')}
{2\kappa_n\left \{ \int_0^au_n^2(r)dr + i u_n^2(a)/2\kappa_n \right \} }\,,
\label{d77}
\end{equation}
which yields the normalization condition for \emph{resonant (quasinormal) states},
\begin{equation}
\int_0^au_n^2(r)dr + i \frac{u_n^2(a)}{2\kappa_n }=1\,.
\label{d78}
\end{equation}
Notice that for bound states, where $\kappa_n=i\eta_n$, with $\eta >0$,  Eq. (\ref{d78}) becomes the usual normalization condition from zero to infinity. For bound states and for a closed system, the \emph{resonance (quasinormal) state} formalism reduces to the usual formalism.

The \emph{Resonant (quasinormal) states} are solutions to the Schr\"odinger equation to the problem,
\begin{equation}
[\kappa_n^2 - H]u_n(r)=0\,,
\label{b1}
\end{equation}
with outgoing boundary conditions:
\begin{equation}
u_n(0)=0, \qquad
\left [\frac{\rm{ d}}{{\rm d} r}u_n(r) \right ]_{r=a}=i\kappa_n u_n(a)\,.
\label{b2}
\end{equation}
The second of the above conditions implies that for $r >a$, $u_n(r)=D_n\exp(i\kappa_n r)$, which leads to complex energy eigenvalues, as first discussed by Gamow \cite{gamow28}, that is, $\kappa_n^2=E_n=\mathcal{E}_n-i\Gamma_n$.

An interesting expression follows by using Green's theorem between equations for $u_n(r)$ and $u_n^*(r)$, and its corresponding boundary conditions, provided  $\alpha_n \neq 0$,
\begin{equation}
\beta_n= \frac{1}{2}\frac{|u_n(a)|^2}{{\displaystyle I_n}}\,,
\label{n1}
\end{equation}
where
\begin{equation}
I_n=\int_0^a |u_n(r)|^2\,dr\,.
\label{b3}
\end{equation}

As discussed in detail in Refs. \cite{gc10,gc11}, the expansion of the outgoing Green's function in terms of the \emph{resonance (quasinormal) states} of the problem may be obtained by considering the integral
\begin{equation}
I=\frac{i}{2\pi} \int_{C} \frac{G^+(r,r';k')}{k'-k}dk'\,,
\label{c1}
\end{equation}
where $C$ corresponds to a  large closed contour of radius $R$ about the origin in the complex momentum $k^\prime$ plane, which excludes all the poles $\kappa_n$  and the real value $k^\prime=k$ located inside, that is, $C=C_S +\sum_n c_n+c_k$. Since  Cauchy's integral theorem establishes that $I=0$, one may use the theorem of residues to evaluate the distinct contours, in view of (\ref{d77}) and (\ref{d78}), to write
\begin{eqnarray}
G^+(r,r';k) &=& \sum_{n=-N}^{N} \frac{u_n(r)u_n(r^{\prime})}{2\kappa_n(k-\kappa_n)}
\nonumber \\ [.3cm]
&&+\frac{i}{2\pi} \int_{C_S} \frac{G^+(r,r';k')}{k'-k}dk'\,.
\label{c2}
\end{eqnarray}
The number of poles appearing in the sum of  (\ref{c2}) may be increased by considering successively larger values of the radius $R$. This follows because the poles are simple and are ordered as $|\kappa_1| \leq |\kappa_2| \leq |\kappa_3| \leq...$ \cite{nussenzveig72}. In the limit as $R \rightarrow \infty$, there will be an infinite number of terms in the sum. In that limit, however,  $G^+(r,r^\prime;k)$ diverges unless $r$ and $r'$ are smaller than the radius $a$ of the interaction potential, or $r=a$ with $r'<a$ and viceversa, but not both of them. We denote the above conditions as $(r,r')^\dag \leq a$. In this case, $G^+(r,r';k') \rightarrow 0$ as $|k| \rightarrow \infty$ along all directions in the complex $k$ plane and hence the integral term in (\ref{c2}) vanishes exactly as shown rigorously in Refs. \cite{gcb79,romo80}. As a result one may write
\begin{equation}
G^+(r,r';k) = \sum_{n=-\infty}^{\infty} \frac{u_n(r)u_n(r^{\prime})}{2\kappa_n(k-\kappa_n)}\,,
\quad  (r,\,r')^{\dagger} \leq \,a.
\label{c3}
\end{equation}
Substitution of (\ref{c3}) into (\ref{d11}) yields, after straightforward manipulations, the closure relationship,
\begin{equation}
\frac{1}{2}\sum_{n=-\infty}^{\infty} u_n(r)u_n(r')=\delta (r-r'); \quad (r,r')^{\dagger} \leq a\,,
\label{c4}
\end{equation}
and the sum rule
\begin{equation}
\frac{1}{2}\sum_{n=-\infty}^{\infty} \frac{u_n(r)u_n(r')}{\kappa_n}=0,  \quad (r,r')^{\dagger} \leq a\,.
\label{c5}
\end{equation}
Noticing that $1/[2\kappa_n(k-\kappa_n)]=1/2k[1/(k-\kappa_n) + 1/\kappa_n]$, one may write (\ref{c3}), in view of (\ref{c5}), as \cite{gc10,gc11}
\begin{equation}
G^+(r,r';k) =\frac{1}{2k} \sum_{n=-\infty}^{\infty} \frac{u_n(r)u_n(r^{\prime})}{k-\kappa_n},
\quad  (r,\,r')^{\dagger} \leq \,a\,.
\label{c6}
\end{equation}
\subsection{The time-dependent solution in terms of resonance (quasinormal) states}

One may obtain the time-dependent solution $\Psi(r,t)$ in term of \emph{resonant (quasinormal states)} by substitution of Eq. (\ref{c6}) into Eq. (\ref{e2}) and the resulting expression into Eq. (\ref{e1}) to  obtain \cite{gc10,gc11},
\begin{equation}
\Psi(r,t)=\sum_{n=-\infty}^{\infty}
\left \{ \begin{array}{cc}
C_nu_n(r)M(y^\circ_n), & \quad  r \leq a \\[.3cm]
C_nu_n(a)M(y_n), & \quad  r \geq a\,,
\end{array}
\label{c7}
\right.
\end{equation}
where
\begin{equation}
C_n=\int_0^{\,a} \Psi(r,0) u_n(r) dr\,,
\label{c8}
\end{equation}
and the functions $M(y_n)$  are defined as \cite{gc10,gc11}
\begin{equation}
M(y_n)=\frac{i}{2\pi}\int_{-\infty}^{\infty}\frac{{\rm e}^{ik(r-a)}{\rm e}^{-ik^2t}}{k-\kappa_n}dk=
\frac{1}{2}{\rm e}^{(imr^2/2 t)} w(iy_n)\,,
\label{c9}
\end{equation}
where $y_n={\rm e}^{-i\pi /4}(1/2t)^{1/2}[(r-a)-2 \kappa_nt]$,  $y_n^{\circ}$ is identical to $y_n$ with $r=a$, and $w(z)=\exp(-z^2)\rm{erfc(-iz)}$ stands for the Faddeyeva or complex error function \cite{abramowitzchap7} for which there exist efficient computational tools \cite{poppe90}.

Assuming that the initial state $\Psi(r,0)$ is normalized to unity, it follows from the closure relationship given by Eq. (\ref{c5}) that,
\begin{equation}
{\rm Re}\,\sum_{n=1}^{\infty} \{C_n\bar{C}_n \}=1\,,
\label{c10}
\end{equation}
where $\bar{C}_n$ is given by
\begin{equation}
{\bar C}_n=\int_0^{\,a} \Psi^*(r,0) u_n(r) dr\,.
\label{c8a}
\end{equation}
Equation (\ref{c10}) shows that the coefficients $C_n$ cannot be interpreted as probability amplitudes, since the sum of their square moduli  does not add up to the norm of $\Psi(r,0)$. Nevertheless, ${\rm Re} \{C_n\bar{C}_n\}$ may be seen to represent the `strength' or `weight' of the initial state in the corresponding \emph{resonance (quasinormal) state} \cite{gcmv07,gcr16}.

Using Eq. (\ref{c7}), one may obtain \emph{resonance (quasinormal) state} expansions of the survival amplitude $A(t)$, and hence of the survival probability $S(t)$, and of the nonescape probability $P(t)$, namely,
\begin{equation}
A(t)= \sum_{n=-\infty}^{\infty}\,C_n\bar{C}_nM(y_n^\circ), \quad S(t)=|A(t)|^2\,,
\label{x1}
\end{equation}
and
\begin{equation}
P(t)=\sum_{n=-\infty}^{\infty}\sum_{l=-\infty}^{\infty}\,C_n C^*_lI_{nl} M(y_n^\circ)
M^*(y_l^\circ)\,,
\label{x2}
\end{equation}
where $I_{nl}=\int_0^a\,u^*_l(r)u_n(r)\,dr$.

In particular, using some properties of the function $M(y_n^\circ)$, the survival amplitude may be written for the exponential and long times regimes as \cite{gcmv07,gc10,gc11},

\begin{eqnarray}
&&A(t) \approx \sum_{n=1}^{\infty}C_n\bar{C}_n {\rm e}^{-i\mathcal{E}_nt} {\rm e}^{-\Gamma_nt/2} \nonumber \\ [.3cm]
&&-i \frac{1}{2(\pi i)^{1/2}} {\rm Im}\left\{\sum_{n=1}^{\infty}\frac{C_n\bar{C}_n}{\kappa_n^3} \right\} \frac{1}{t^{\,3/2}}\,,
\label{x3}
\end{eqnarray}
or to discuss the ultimate fate of a decaying quantum state \cite{gcmv13}.

Equations (\ref{x1}), (\ref{x2}), and (\ref{x3}) should be contrasted with the `black-box' type of calculations that provide  Eqs. (\ref{11d}) and (\ref{12d}) in terms of \emph{continuum wave functions}. However, as pointed out above, both formulations give identical numerical results.

It is worth mentioning that the formalism outlined above differs from the so called rigged Hilbert space formulation in many respects, as discussed in Refs. \cite{mgcm05,gc10}. For example, since in that approach the poles located on the third quadrant of the $k$ plane are not taken explicitly into consideration, there is no analytical description as that given by Eqs. (\ref{x1}), (\ref{x2}) and  Eq. (\ref{x3}).

\section{Resonance (quasinormal) expansion of $|C(k)|^2$}

Substitution of  Eq. (\ref{c6}) into Eq. (\ref{n2}) yields the expansion of the \emph{continuum wave function} $\psi^+(k,r)$ in terms of \emph{resonant (quasinormal) states},
\begin{equation}
\psi^{+}(k,r)=-\sqrt{\frac{2}{\pi}}\,\frac{1}{2}{\rm e}^{-ika} \sum_{n=-\infty}^{\infty} \frac{u_n(a)u_n(r)}{k-\kappa_n}, \quad r<a\,.
\label{c11}
\end{equation}
We may then substitute (\ref{c11}) into the expression of $C(k)$ given by (\ref{d35}), using (\ref{c8}), to obtain
\begin{equation}
C(k)=-\left [\sqrt{\frac{2}{\pi}}\,\frac{1}{2}{ \rm e}^{-ika} \sum_{n=-\infty}^{\infty} \frac{{\bar C}_nu_n(a)}{k-\kappa_n}\right ]^*\,.
\label{c12}
\end{equation}
One may run the above sum from $n=1$ up to infinity, by noticing that $\kappa_{-n}=-\kappa^*_n$ and $u_{-n}(r)=u^*_n(r)$ \cite{gc10,gc11}. Hence, this allow us to write the expression for $|C(k)|^2$ as,
\begin{eqnarray}
&&|C(k)|^2=\frac{1}{\pi}\sum_{n=1}^{\infty} |C_n|^2I_n \frac{\beta_n}{(k-\alpha_n)^2+\beta_n^2}
\nonumber \\ [.3cm]+
&&\frac{1}{\pi}\sum_{n=1}^{\infty} |C_n|^2I_n \frac{\beta_n}{(k+\alpha_n)^2+\beta_n^2}
\nonumber \\ [.3cm]
&& +\frac{1}{\pi} {\rm Re}\left \{\sum_{n \pm s}^\infty \frac{C_nC_su_n(a)u_s(a)}{[(k-\alpha_n)+i\beta_n][(k+\alpha_s)-i\beta_s]}\right \}\,,
\label{c13}
\end{eqnarray}
where we have used Eqs. (\ref{n1}) and (\ref{b3}). Using (\ref{c13}) allows us also to calculate
\begin{eqnarray}
&&\int_0^{\infty}|C(k)|^2\,dk=\sum_{n=1}^{\infty} |C_n|^2\,\int_0^a |u_n(r)|^2\,dr \nonumber \\ [.3cm]
&&-\,{\rm Re} \left \{\sum_{n\pm s}^{\infty} C_nC_s \,i\,\frac{u_n(a)u_s(a)}{\kappa_n+\kappa_s}\right \}=1\,.
\label{c14}
\end{eqnarray}
It is worth mentioning that Eq. (\ref{c13}) for $|C(k)|^2$, is given by a sum of resonance peaks having a Lorentzian shape that depends on the resonance terms $\alpha_n$ and $\beta_n$, each peak  multiplied by the coefficients $|C_n|^2$, formed by the overlap of the initial state $\Psi(r,0)$ and the corresponding resonance  (quasinormal) state $u_n(r)$ and $I_n$, given by the integral of $|u_n(r)|^2$ along the internal interaction region plus an interference term. One sees that the contribution of each resonance peak depends on the value attained by the corresponding product $|C_n|^{2}I_n$, which in view of Eq. (\ref{c14}) does not add up to a unity value, precisely due to the contribution of the interference term. Hence, in general,
\begin{equation}
\sum_{n=1}^{\infty} |C_n|^2 I_n > 1\,.
\label{c15}
\end{equation}

The role of the initial state $\Psi(r,0)$ is crucial. In the case of an initial state that overlaps strongly with one of the \emph{resonance (quasinormal) states} of the system, say, $u_r(r)$, as in the example below, one may see that ${\rm Re}\,C_r\bar{C}_r \approx 1$, and also $|C_r|^2I_r \approx 1$. In that case Eq.  (\ref{c14}) may be written as
\begin{equation}
\int_0^{\infty}|C(k)|^2\,dk \approx  |C_r|^2I_r \approx 1\,.
\label{c16}
\end{equation}
In general, however, Eq. (\ref{c14}) suggests that the coefficients $|C_n|^2$ and $I_n$ may possess a quasi-probabilistic nature \cite{halliwell13}.

It is worth commenting that it has been a common practice in the literature to approximate $|C(k)|^2$  by just a single Lorentzian, which therefore implies that the coefficient involving the initial state has a unity value, as in the work by Khalfin, who showed that the exponential decay law cannot hold at all times \cite{khalfin58}. In general, however, this is not justified.  When more resonance levels are involved in  the decay process and the initial state overlaps with several resonance (quasinormal) levels, a more complex decaying behavior arises \cite{gcrv07,gcrv09,cgcrv11}.

\begin{figure}[!tbp]
\includegraphics[width=3.3in]{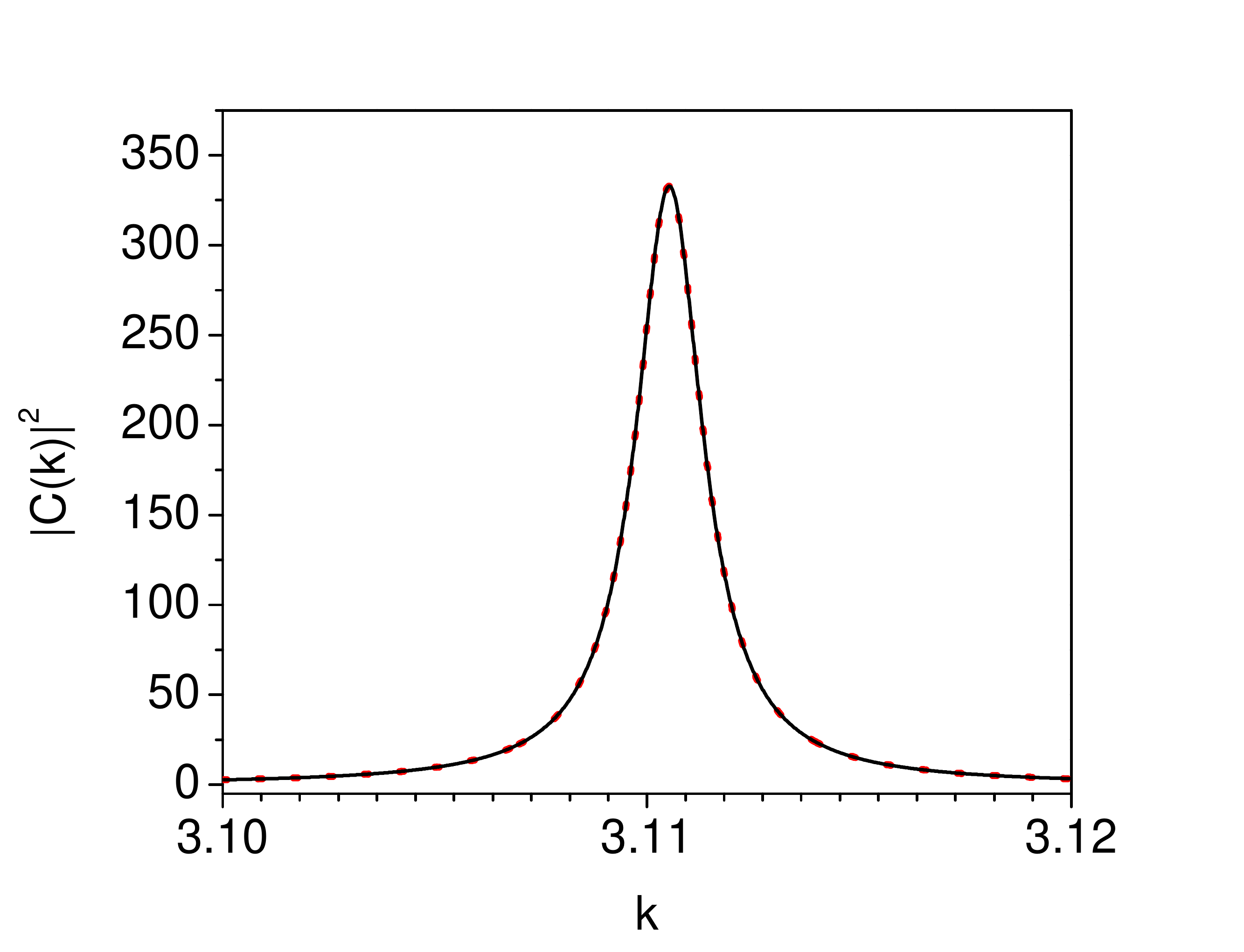}
\caption{Plot of $|C(k)|^2$ using \emph{continuum wave functions} (full line) and \emph{resonance (quasinormal) states} (dotted line) versus $k$ for the $delta$-shell potential with parameters $\lambda=100$ and $a=1$. See text.}
\label{figure1}
\end{figure}
\section{Model}

In order to illustrate our findings, we consider the exactly solvable model given by a $\delta$-shell potential of intensity $\lambda$ and radius $a$, for zero angular momentum,
\begin{equation}
V(r)=\lambda \delta(r-a)
\label{m1},
\end{equation}
and an initial state, the infinite box state,
\begin{equation}
\Psi(r,0)= \left (\frac{2}{a} \right )^{1/2} \sin \left (\frac{\pi r}{a}\right )\,.
\label{m2}
\end{equation}
This model has also been used in Ref. \cite{gcmv12} to illustrate that the formulations or the probability density $\Psi(r,t)$ in terms of \emph{continuum wave functions} and \emph{resonance (quasinormal) states} yield identical results for the time evolution of decay.

Since the initial state given by (\ref{m2}) is confined within the interaction region, the \emph{continuum wave function} reads, using Eq.  (\ref{e4a}),
\begin{equation}
\psi^+(k,r) = \sqrt{\frac{2}{\pi}}\, \frac{\sin(kr)}{J_+(k)}, \quad r < a\,,
\label{m3}
\end{equation}
where the Jost function $J_{+}(k)$ reads,
\begin{equation}
J_+(k)=2ik + \lambda ({\rm e}^{2ik a}-1)\,.
\label{m4}
\end{equation}
Using Eqs. (\ref{m2}), (\ref{m3}) and (\ref{m4}) into Eq. (\ref{d35}) yields
\begin{equation}
C(k)=\frac{2ik}{\sqrt{\pi a}}\,\,\frac{1}{J_+(k)}\,
\left (\frac{\sin[(s-k)a]}{s-k}-\frac{\sin[(s+k)a]}{s+k}\right )\,,
\label{m5}
\end{equation}
with $s=\pi/a$, from which one may calculate  $|C(k)|^2$.
\begin{figure}[!htbp]
\includegraphics[width=3.3in]{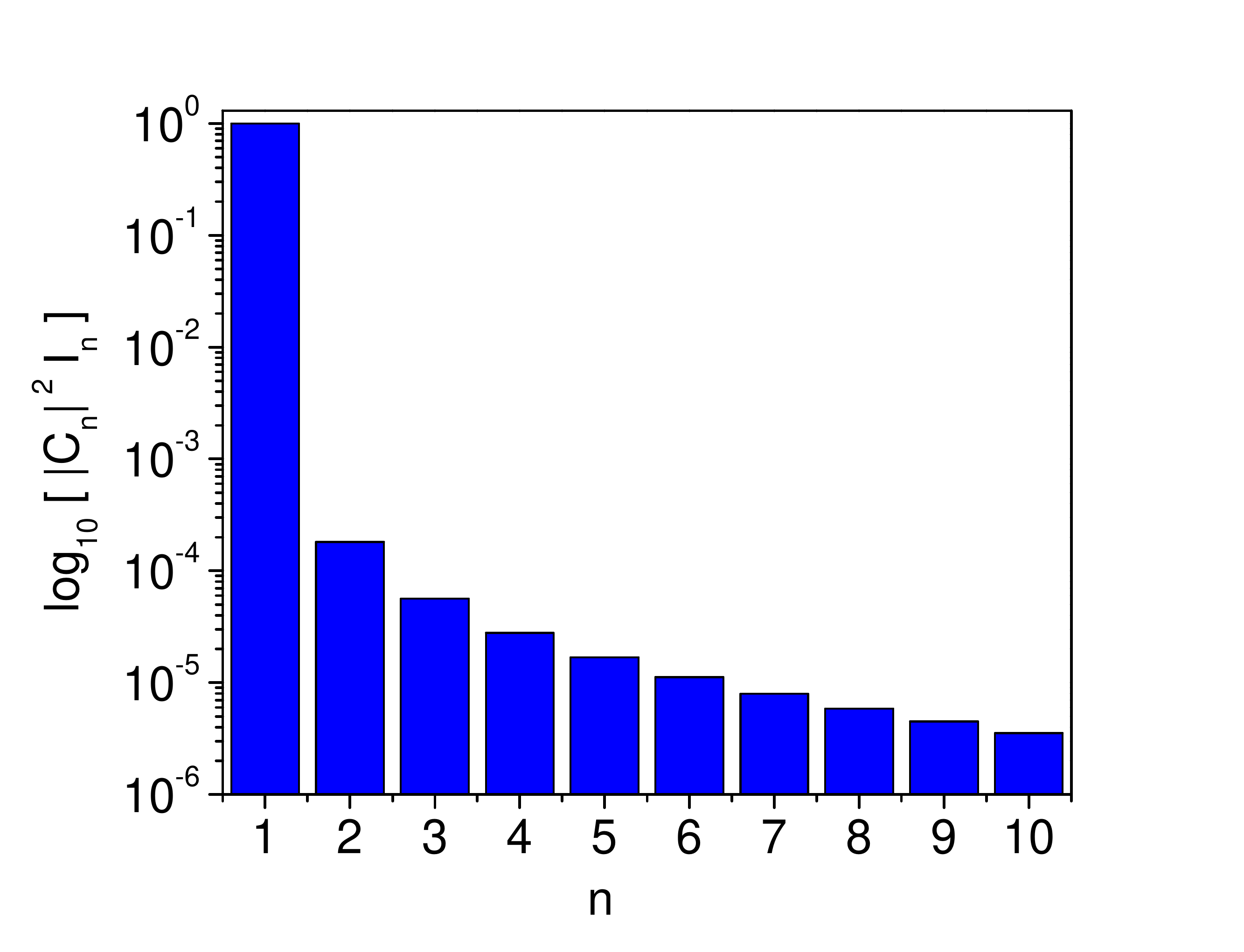}
\caption{Plot of $log_{10}$ $|C_n|^2I_n$ as a function of the resonance (quasinormal) states $n$ used in the calculation of Fig. \ref{figure1}. See text.}
\label{figure2}
\end{figure}

Similarly, the \emph{resonance (quasinormal) states} along the internal interaction region are given by
\begin{equation}
u_n(r)= A_n \,\sin (\kappa_nr), \quad r\leq a\,,
\label{m6}
\end{equation}
where, using Eq. (\ref{d78}), the normalization $A_n$ reads,
\begin{equation}
A_n=\left [\frac{2\lambda }{\lambda a+ {\rm e}^{-2i \kappa_n a}} \right ]^{1/2}\,.
\label{m7}
\end{equation}
The set of complex poles $\{\kappa_n\}$ follows from the zeros of the Jost function given by Eq. (\ref{m4}), namely,
\begin{equation}
J_+(\kappa_n)=2i\kappa_n + \lambda ({\rm e}^{2i\kappa_n a}-1)=0\,.
\label{m8}
\end{equation}
The solutions to the above equation has been discussed elsewhere \cite{gc10}. For example, for $\lambda \gg 1$,
they admit the approximate analytical solution,
\begin{equation}
\kappa_n \approx \frac{n\pi}{a} \left (1-\frac{1}{\lambda a}\right ) -
i \,\frac{1}{a}\left( \frac{n\pi}{\lambda a}\right )^2\,.
\label{m9}
\end{equation}
One may then use iterative methods as the Newton-Raphson method to get the solution with the desired degree of approximation.

Using Eqs. (\ref{m2}), (\ref{m6}), and (\ref{m7}) into Eq. (\ref{c8}) yields,
\begin{equation}
C_n=\frac{A_n}{\sqrt{2a}}
\left (\frac{\sin[(s-\kappa_n)a]}{s-\kappa_n}-\frac{\sin[(s+\kappa_n)a]}{s+\kappa_n}\right )\,,
\label{m10}
\end{equation}
with $s=\pi/a$, from which $|C_n|^2$ can be calculated. In a similar fashion, using (\ref{m6}) and (\ref{m7}) into Eq. (\ref{b3}) one obtains,
\begin{equation}
I_n= \frac{|A_n|^2}{4}\left[ \frac{\sinh(2\beta_n a)}{\beta_n} -\frac{\sin(2\alpha_n a)}{\alpha_n}\right ]\,.
\label{m11}
\end{equation}

Figure \ref{figure1} provides a plot of the coefficient $|C(k)|^2$ as a function of $k$ around the first resonance level $\kappa_1=\alpha_1-i \beta_1$ of the $delta$-shell potential with intensity $\lambda=100$ and radius $a=1$. The coefficient $|C(k)|^2$ is evaluated using the basis of \emph{continuum wave functions} given by Eq. (\ref{m5}) (solid line) and by using the basis of \emph{resonance (quasinormal) states} corresponding to Eq. (\ref{c12}), which is identical to Eq. (\ref{c13}), with $C_n$ given by Eq. (\ref{m10}). One sees that they yield identical results. In this case, as shown in Fig. \ref{figure2} which exhibits a plot of $\log_{10}\, |C_n|^2I_n$ for the first $10$ resonance levels, the term $n=1$ dominates, and hence this term is sufficient to obtain an excellent description of $|C(k)|^2$. Our calculations show, for this and other cases, the $I_n \gtrapprox 1$, and therefore this
indicate that the coefficient $|C_n|^2$ is the main ingredient to determine the value of a given resonance term to the probability $|C(k)|^2$. Figure \ref{figure2} shows also that the value of $|C(k)|^2$ for a value of $k$ around any of the other resonance values of the system is very small.

\section{Concluding remarks}

Equations (\ref{c13}) and (\ref{c14}) constitute  the main result of this work. They refer to a non-Hermitian analytical formulation  that lies outside the conventional Hilbert space, which however yields identical numerical results as a formulation based on \emph{continumm wave functions}. The distinct resonance (quasinormal) contributions, given by Eq. (\ref{c13}), provide a deeper insight in the way that $|C(k)|^2$ attains a given value. In particular, the role of the coefficients  $|C_n|^2$ which involve the overlap of the initial state with the corresponding  \emph{resonance (quasinormal) state}.  However, as shown by Eq. (\ref{c14}), the sum over the coefficients $|C_n|^2I_n$ does not add up to unity, and hence in general they cannot be interpreted as probabilities, although they may be seen to represent the `strength' or `weight' of the initial state in the corresponding \emph{resonance (quasinormal) state}. Presumably, they might be considered as quasi-probabilities \cite{halliwell13}, but this requires further study. Our formulation clearly shows the relevant role played by initial states. The recent developments on artificial quantum systems may lead to their control and manipulation \cite{jochim11}. Our results might be of interest in the recent discussions on the reality of the wave function.

We would like to end by quoting  Shakespeare's Hamlet: \emph{There are more things in Heaven and Earth, Horatio, than are dreamt of in your philosophy}.

\begin{acknowledgements}
G.G-C. acknowledges the partial financial support of DGAPA-UNAM-PAPIIT IN105216, Mexico.
\end{acknowledgements}
%

%\bibliography{refs_sergio}

%
\end{document}